\documentclass[aps,showpacs,prb,twocolumn]{revtex4}

\usepackage{amssymb}
\usepackage{amsmath}
\usepackage{bm}
\usepackage{graphics}
\usepackage{natbib}

\begin{document}
\title{Spin-bias driven electron properties of a triple-quantum-dot ring}
\author{Weijiang Gong$^{1,3}$}
\author{Xuefeng Xie$^{1}$}
\author{Yu Han$^{1,2}$}
\author{Guozhu Wei$^{1,3}$}\email[Author to whom correspondence should be addressed. Electronic mail: ]
{guozhuwei02@sina.com} \affiliation{ 1. College of Sciences,
Northeastern University, Shenyang 110004, China\\
2. Department of Physics, Liaoning University, Shenyang 110036, China \\
3. International Center for Material Physics, Acadmia Sinica,
Shenyang 110015, China}

\date{\today}

\begin{abstract}
Electron transport through a three-electrode triple-quantum-dot ring
with the source electrode of spin-dependent splitting of chemical
potentials (spin bias) is theoretically investigated. We find clear
charge and spin currents in the drain electrodes driven by the spin
bias, despite the absence of charge bias between the source and
drain electrodes, and their directions and amplitudes can be
adjusted by altering the structure parameters or magnetic field. The
distinct characteristics of spin-bias driven persistent charge and
spin currents in the ring are also shown. When an appropriate charge
bias is applied, the single-spin electron motion can be achieved in
this structure.
\end{abstract}
\maketitle

\bigskip
The manipulation and control of the behaviors of electron spins in
nanostructures have become one subject of intense investigation due
to its relevance to quantum computation and quantum
information.\cite{Ohno,Dasarma} The electron spin in quantum dot
(QD) is a natural candidate for the qubit, QD has therefore been
regarded as an elementary cell of such a field and much attention
has been paid to the manipulation of the electron spin degree of
freedom in QD for its application.\cite{Loss,Loss2} However, one of
the challenge is to efficiently realize the spin transport without
any appreciable charge transfer, for the reason that the spin is
difficult to be manipulated by a charge bias voltage. Many schemes
have been proposed to work out this problem, based on the case of a
charge bias between two leads with a magnetic field or the spin-obit
coupling for a QD system.\cite{Datta,Rashba,Rashba2,Sun} Despite
these existed works, any new suggestions to realize the pure spin
current are still necessary. Recently, it has been reported that
spin bias in leads for mesoscopic systems can be
feasible,\cite{Nagaosa,Hubner,Cui,Li,Jpn} which induces rich
physical phenomena and potential applications different from the
traditional charge bias.

\par

In this letter, we propose a theoretical approach to realize the
pure spin transport in a three-terminal triple-QD ring by assuming
the source electrode of spin bias. Our theoretical investigation
indicates that it is possible to form the pure spin current in
either normal lead. Meanwhile, the spin-bias driven pure persistent
spin current is apparent with its adjustable direction and
amplitude. In the case of an appropriate charge bias applied, the
single-spin electron transmission can also be achieved.
\par
\begin{figure}
\centering \scalebox{0.35}{\includegraphics{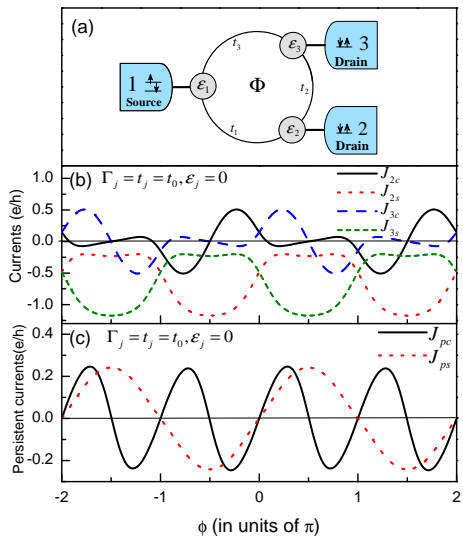}} \caption{ (a)
Schematic of a three-terminal triple-QD ring structure with spin
bias in the source electrode. The currents vs the magnetic phase
factor $\phi$ are shown in (b).  (c) The persistent currents vs
$\phi$. The parameter values are $\Gamma_j=t_j=t_0$ and
$\varepsilon_j=0$. \label{structure}}
\end{figure}

The structure under consideration is illustrated in
Fig.\ref{structure}(a), and the Hamiltonian to describe the electron
motion in the system can be written as ${\cal
H}=\sum_{\sigma,j,k}\varepsilon _{jk\sigma}c_{jk\sigma}^\dag
c_{jk\sigma }+\sum_{j, \sigma}\varepsilon
_{j}d^\dag_{j\sigma}d_{j\sigma}+\sum_j U_j
n_{j\uparrow}n_{j\downarrow}
+\sum_{\sigma,l=1}^{2}t_{l}e^{i\phi/3}d^\dag_{l+1\sigma}d_{l\sigma}
+t_{3}e^{i\phi/3}d^\dag_{1\sigma}d_{3\sigma}+\underset{\sigma
j,k}{\sum }V_{j} d^\dag_{j\sigma}c_{jk\sigma}+\mathrm {H.c.}$. Here
$c_{jk\sigma}^\dag$ and $d^\dag_{j\sigma}$ ( $c_{jk\sigma}$ and
$d_{j\sigma}$ ) is an operator to create (annihilate) an electron of
the continuous state $|k,\rangle$ in lead-$j$ (QD-$j$) with $\sigma$
being the spin index, and $\varepsilon _{jk\sigma}$ and $\varepsilon
_{j}$ is the corresponding single-particle energy with $j=1$ to $3$.
$U_j$ denotes the intradot Coulomb interaction, and $t_j$ represents
the interdot hopping coefficient. In the case of the identical QDs
distributed in the ring equidistantly, we can use uniform parameters
$t_0$ to denote the interdot hopping coefficients, which is real at
the zero-magnetic-field case. If a magnetic field is applied
perpendicular to the ring plane, in the symmetrical gauge it becomes
$t_j=t_0e^{i\phi/3}$ with $\phi=2\pi\Phi/\Phi_0$, where $\Phi$ is
the magnetic flux threading the ring and $\Phi_0=h/e$ is the
magnetic flux quantum. By means of the Green function technique, at
zero temperature, the current flow in lead-$j$ can be written as $
J_{j\sigma}=\frac{e}{h}\sum_{j'}\int d\omega
T_{jj'\sigma}(\omega)[f_{j\sigma}(\omega)-f_{j'\sigma}(\omega)]
$,\cite{Meir,Gong1} where $T_{jj',\sigma}(\omega)=\Gamma_j
 G^r_{jj',\sigma}(\omega)\Gamma_{j'}G^a_{j'j,\sigma}(\omega)$
is the transmission function, describing the electron tunneling
ability from lead-$j$ to lead-$j'$, and $f_{j\sigma}(\omega)$ is the
Fermi distribution function of $\sigma$-spin electron in lead-$j$.
$\Gamma_j=2\pi |V_{j}|^2\rho_j(\omega)$, the strength of the
coupling between QD-$j$ and lead-$j$, can be usually regarded as a
constant. $G^r$ and $G^a$, the retarded and advanced Green
functions, obey the relationship $[G^r]=[G^a]^\dag$. By adopting the
equation of motion method, we can work out the retarded Green
function by means of the second-order (i.e., Hubbard) approximation
to truncate the higher-order Green function without the electron
correlation being considered.\cite{Gong1} In order to clarify the
electron and spin transport behaviors in such a structure, the
charge and spin currents in lead-$j$ are defined respectively as
$J_{jc}=J_{j\uparrow}+J_{j\downarrow}$ and
$J_{js}=J_{j\uparrow}-J_{j\downarrow}$. In addition, the persistent
current for the $\sigma$-spin electron in the QD ring can be given
by $J_{p\sigma}=\int d\omega
\mathrm{Re}[t_lG^<_{ll+1,\sigma}(\omega)]$ \cite{Hancock} with
$G^<_{\sigma}(\omega)=G^r_{\sigma}(\omega)\Sigma^<_{\sigma}G^a_{\sigma}(\omega)$
and $\Sigma^<_{\sigma}=\sum_j i\Gamma_jf_{j\sigma}(\omega)$, so the
persistent charge and spin currents in such a ring are thereby
well-defined with $J_{pc}=J_{p\uparrow}+J_{p\downarrow}$ and
$J_{ps}=J_{p\uparrow}-J_{p\downarrow}$.
\par
Consider the chemical potentials of electrons in lead-1 ( the Source
) are spin-dependent, i.e., $\mu_{1\sigma}=\varepsilon_F+{e\over
2}(V_c+\sigma V_s)$, whereas in the other leads ( the Drains )
$\mu_{2\sigma}=\mu_{3\sigma}=\varepsilon_F-{e\over 2}V_c$ with $V_c$
being the charge bias voltage between the leads and $V_s$ the spin
bias, we now proceed on to investigate the electron transport
properties in such a double-channel structure. Before calculation,
we assume the uniform interdot hopping $t_0$ as the unit of energy
and $\varepsilon_F$ as the zero point of this system, and, the spin
bias is taken to be $eV_s=2t_0$.
\par

With the adjustment of the threading magnetic flux, the current
flows in the drains (i.e., lead-2 and lead-3) are first calculated
in the zero-charge-bias case. The relevant parameters take the
values as $\Gamma_{j}=t_0$, whereas $\varepsilon_j$, the QD levels,
are fixed at zero. The many-body terms are here not taken into
account, since we are only interested in the electron and spin
transport. Fig.\ref{structure}(b) shows the changes of the charge
and spin currents in the drains versus the magnetic phase factor
$\phi$, respectively. It can be first found that, the amplitudes and
directions of the charge currents in the two drains are tightly
dependent on the tuning of magnetic flux with the periodicity
$2\pi$, but obviously they vary out of phase; on the other hand, due
to the consideration of spin bias, there also distinctly emerge spin
currents in the two drains, and they reach their maxima alternately
with the adjustment of magnetic flux. Furthermore, \emph{in the
range of the magnetic flux factor $\phi$ from $2n\pi$ to $(2n+1)\pi$
[ $(2n-1)\pi$ to $2n\pi$ ]}, both the amplitudes of the charge and
spin currents in lead-2 (lead-3) are suppressed, which indicates
that the electron traveling between lead-1 and lead-2 (lead-3) is
forbidden. In contrast to the above case, when the magnetic flux is
tuned in the other regimes the electron transmission in the two
channels are correspondingly allowed and \emph{at the position of
$\phi=(2n-{1\over2})\pi$ [ $\phi=(2n+{1\over2})\pi$ ]} the direction
of the charge current is just inverted, where the pure spin current
comes into being in lead-2 ( lead-3 ) with its maximal amplitude.
Also, we can find that in the regimes where the electron transport
is allowed, the amplitudes of the spin currents are larger than
those of the charge currents, so high-efficiency spin injection can
be realized in this structure.
\par

Since the structure of quantum ring, the persistent current is
another concern. In Fig.\ref{structure}(c), both the persistent
charge and spin currents versus $\phi$ are shown. Clearly, by the
adjustment of magnetic flux the oscillation of the persistent charge
and spin currents with different periodicities. In the vicinity of
$\phi=(n+{1\over 2})\pi$, namely, the magnetic phase factor is the
odd multiple of $\pi\over 2$, $J_{ps}$ reaches its maximum, where
the persistent charge current $J_{pc}$ is just at a zero point.
Moreover, with the variation of magnetic flux from $\phi=(2n-{1\over
2})\pi$ to $(2n+{1\over 2})\pi$ the polarization direction of the
persistent spin current in inverted. Thus, with the help of these
results, one can understand that in this structure, the spin bias
can drive the appearance of the pure persistent spin currents, the
polarization direction of which can be changed by tuning the applied
magnetic field.
\par
\begin{figure}
\centering \scalebox{0.35}{\includegraphics{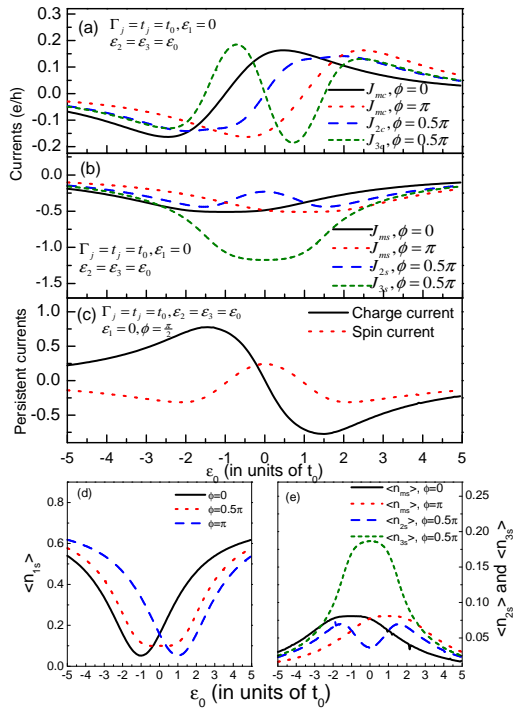}} \caption{ (a)
and (b) The currents vs the QD levels $\varepsilon_0$ with the phase
factor $\phi=0$, $0.5\pi$, and $\pi$. (c) The persistent currents vs
$\varepsilon_0$ in the case of $\phi=0.5\pi$. (d) and (e) The spin
accumulations in the respective QDs in the cases of $\phi=0$,
$0.5\pi$, and $\pi$, respectively. \label{Cond}}
\end{figure}

One of the well-known characteristics of QD is its tunable level via
the adjustment of gate voltage in experiment. So, it is necessary
for us to pay attention to the behaviors of the currents with the
change of QD levels. In order to especially observe the spin
accumulation of QD-1, $\varepsilon_1$ is fixed at zero, but the
levels of the other QDs are taken to be $\varepsilon_0$, which can
be shifted by tuning the gate voltage. The profiles of the relevant
quantities as functions of $\varepsilon_0$ are plotted, and the
corresponding results are exhibited in Fig.\ref{Cond}, where, for
simplicity, the many-body terms are ignored. We can readily find, in
Fig.\ref{Cond}(a) that, similar to the results in Fig.1, for the
case of $\phi=n\pi$ the charge currents in the two drains are the
same as each other. But the shift of gate voltage can effectively
change the directions of the charge currents, namely, in the absence
of magnetic flux when $\varepsilon_0$ exceeds the position of
$\varepsilon_0=-t_0$, the direction of the charge currents in the
drains will be inverted, whereas when $\phi=\pi$ such a point is
shifted to the position of $\varepsilon_0=t_0$. However, when the
magnetic flux is tuned to $\phi={\pi\over 2}$, the profiles of the
charge currents are separate from each other with their remarkable
differences, though both of them are asymmetric about the point of
$\varepsilon_0=0$. In Fig.\ref{Cond}(b) we show the spectra of the
spin currents in the drains, as a result, it is seen that the
symmetric points of the spin currents just correspond to the
anti-symmetric points of the charge currents and at these positions
the spin currents present their exterma, respectively. Consider the
case of $\phi={\pi\over 2}$, the spectra of persistent charge and
spin currents are shown in Fig.\ref{Cond}(c). It is clear that when
$\varepsilon_0$ goes beyond the zero point of the energy the
direction of the persistent charge current is inverted, but with
respect to the persistent spin current, its spectrum is just
symmetric about the position of $\varepsilon_0=0$ where the maximum
of its amplitude emerges. In Fig.\ref{Cond}(d) and
Fig.\ref{Cond}(e), we turn our attention to the spin accumulation in
respective QDs. Associated with the results of the spin currents in
Fig.\ref{Cond}(b), it can be found that where there occurs the
active spin transport are just the positions of the minimal spin
accumulation in QD-1 in despite of its fixed level, but meanwhile in
such regions in the other QDs the spin accumulations reaches their
maxima (Here $\langle n_{js}\rangle=\langle
n_{j\uparrow}\rangle-\langle n_{j\downarrow}\rangle$ is considered
to denote the spin accumulations in the respective QDs).
\par
\begin{figure}
\centering \scalebox{0.35}{\includegraphics{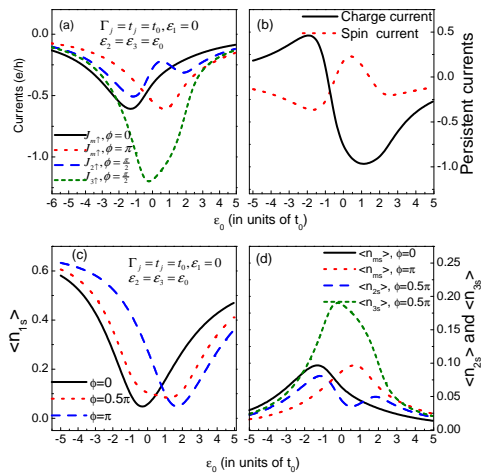}} \caption{ The
case of finite charge bias between the source and drains with
$eV_c=t_0$. (a) and (b) The currents vs the QD levels
$\varepsilon_0$ with $\phi=0$, $0.5\pi$, and $\pi$. (c) The
persistent currents vs $\varepsilon_0$ in the case of $\phi=0.5\pi$.
(d) and (e) The spin accumulations in the respective QDs in the
cases of $\phi=0$, $0.5\pi$, and $\pi$, respectively. \label{Cond2}}
\end{figure}

It is interesting that when a charge bias in introduced with
$eV_c=t_0$, only the spin-up electron in the source can flow through
this structure to the drains, and in such a case the spin current is
just the charge current. Accordingly, in Fig.\ref{Cond2} we show the
spin-up electron properties driven by the charge and spin bias. It
is seen that only the amplitude of the current can be adjusted by
the tuning of magnetic flux or the shift of gate voltage, and in the
case of $\phi={\pi\over 2}$ the spin polarization in lead-3 is more
apparent than that in lead-2. On the other hand, the application of
charge bias influences the persistent currents in a nontrivial way.
In such a case, the persistent charge current is apt to rotate along
the negative direction, because of the left shift of the position
where the direction of the persistent charge current inverts and the
increase of the amplitude of the persistent charge current in the
negative direction. This brings about the reversal of the direction
of the pure persistent spin current, in comparison with that in the
case of zero charge bias. Simultaneously, as shown in
Fig.\ref{Cond2}(c) and (d), the extrema of the spin accumulations in
respective QDs well coincide with the results in Fig.\ref{Cond2}(a).
\par
In conclusion, we have discussed the electron transport through a
triple-quantum-dot ring of three terminals. We found that despite
the absence of the charge bias between the electrodes, in the drain
electrodes there appear distinct charge and spin currents driven by
the spin bias in the source electrode. Besides, in such a QD ring
the persistent charge and spin currents driven by the spin bias are
also remarkable and the directions and amplitudes of them can be
adjusted by altering the structure parameters. On the other hand,
the application of a finite charge bias between the source and
drains of this structure can lead to the single-spin electron
transmission. With respect to the many-body terms, we have to point
out that the influence of the Coulomb repulsion on the electron
transport is to divide the electron transport spectra into two
groups, and in each group the results are similar to that in the
noninteracting case. However, based on our calculated results, the
amplitudes of the currents are correspondingly suppressed compared
with those in the case of zero electron interactions, which can be
clarified by the fact that the Coulomb repulsion reflects the
localization of electrons.

\clearpage

\bigskip

\end{document}